
\input jnl
\def\cor#1{\langle{#1}\rangle}

\def\mag#1{\vert #1\vert}
\dateline
\title Randomly Branched Polymers and Conformal Invariance
\author Jeffrey D. Miller
Service de Physique Theorique de Saclay
91191 Gif-sur-Yvette Cedex France
\author Keith De'Bell
Trent University
Department of Physics
Peterborough,
Ontario, Canada K9J 7B8

\abstract We argue
that the field theory that descibes randomly
branched polymers is not generally conformally invariant
in two dimensions at its critical point.
In particular, we show
(i) that the most natural formulation of conformal invariance
for randomly branched polymers leads to inconsistencies;
(ii) that the free field theory obtained by setting
the potential equal to zero in the branched polymer
field theory is not even classically conformally
invariant; and (iii) that numerical
enumerations of the exponent $\theta (\alpha )$,
defined by
$T_N(\alpha )\sim \lambda^NN^{-\theta (\alpha ) +1}$,
where $T_N(\alpha )$ is number of distinct configuratations of
a branched polymer rooted near the apex of
a cone with apex angel $\alpha$, indicate that
$\theta (\alpha )$ is not
linear in $1/\alpha$ contrary to what conformal invariance
leads one to expect.

\vfill
\noindent SPhT/92-145

\endtitlepage

{\bf I. Introduction}

It is widely believed that statistical mechanical systems
at the critical point of a second order phase transition
are conformally invariant on scales much larger than
any microscopic distance\refto{polyakov}.\footnote{$^a$}{
More precisely, translational, rotational, and scale
invariance, in systems with only short range interactions,
is thought to imply conformal invariance.}
In two dimensions, the conformal algebra is infinite
dimensional and conformal invariance places strong
constraints on the fixed point correlation functions.
The analysis of these constraints, starting with the
paper of Belavin, Polyakov, and Zamolodchikov\refto{belavin}, has lead
to a rather complete understanding of two dimensional critical
theories\refto{cardy}.
The conformal properties of most
random geometrical systems:
linear polymers, branched
polymers with fixed topology, theta polymers, percolation, dense polymers,
etc. in two dimensions are well understood\refto
{duplantier}.
Randomly branched polymers\refto{lubensky} are a notable exception.
Numerical enumerations of randomly branched polymers
in the wedge\refto{DBL85}, and transfer matrix calculations of
randomly branched polymers
in the strip\refto{derrida}, yield results at variance with the
naive predictions of conformal invariance.

In this paper we argue that the critical point
of the theory describing
randomly branched polymers
is not generally conformally invariant
in two dimensions.
The arguments are quite simple.
We state them here in outline.

(i) We first show that the most natural formulation
of conformal invariance for the field
theory that describes randomly branched polymers\refto{lubensky,
parisi}
leads to inconsistencies.
We proceed by first
assuming the field theory in question to be conformally
invariant, and then by studying the consequences of this assumption.

In order to study the consequences of conformal
invariance, we have first to say
how the fields in the randomly branched polymer field
theory transform under conformal transformations.
This leads us to ``the most natural formulation''
of conformal invariance for randomly branched polymers.
The most natural formulation of conformal invariance
follows immediately from the
supposition that the most relevant fields
in the randomly branched polymer problem
are those which appear in the effective Hamiltonian
that defines the field theory (see equation 1 below).
This is a very mild requirement;
the field theory
in question describes the large distance physics of
randomly branched polymers, so we expect that
the most relevant operators in the branched polymer problem
to be those of the field theory, i.e. those which
appear in the effective Hamiltonian, and
which are integrated over in the functional integrals.

Given, as seems likely, that the most relevant
operators in the randomly branched polymer problem
are those appearing  in the effective
Hamiltonian, it follows that the most relevant field in the
effective Hamiltonian,
the field with lowest scaling dimension, is primary in the
sense of conformal field theory (that is to say,
$\it{if}$ the theory is conformally invariant, $\it{then}$
 the lowest dimension
field must be primary).
For if this field were not primary, it
would, by the usual structure of conformal field
theories, have to be a secondary field, and therefore the descendent
of a field with smaller scaling dimension, and
by assumption there is no such field.

Given that the lowest dimension field in the
effective Hamiltonian is primary, it is staightforward
to show, by conformally mapping the plane to the cone, that
one obtains inconsistent predictions for the
exponent $\theta (\alpha )$ defined by
$T_N(\alpha )\sim \lambda^NN^{-\theta (\alpha ) +1}$,
where $T_N(\alpha )$ is number of distinct configurations of
a branched polymer rooted near the apex of
a cone with apex angle $\alpha$.
{}From this we can conclude that either the
most relevant operators in the branched polymer
theory do not appear in the effective Hamiltonian
(which, we repeat, seems to us unlikely),
or that the field theory is not
conformally invariant.

(ii) The field theory that describes
randomly branched polymers has
a Parisi-Sourlas supersymmetry
\refto{parisi}.
To further investigate the question of the conformal
invariance of randomly branched polymers,
we consider the simplest theory that has the
same kind of Parisi-Sourlas supersymmetry.
This simple theory is obtained by setting
the potential equal to zero in the branched polymer
effective Hamiltonian.
We show that this free
field theory is not even classically conformally invariant.
The reason for this is that the free Parisi-Sourlas theory
is essentially
a higher derivative Gaussian theory,
and it is known that, in higher derivative theories,
scale invariance does not imply conformal invariance
\refto{zinn}.
The lack of conformal invariance in the free theory
does not of itself imply that the
interacting theory of interest is not conformally
invariant; it does however make it seem
less likely.

(iii) Finally, we present numerical enumerations of
of randomly branched polymers in the cone.
If randomly branched polymers were conformally
invariant, the exponent $\theta (\alpha )$ in the
cone would be simply related to the same
exponent in the plane. More precisely,
conformal invariance leads one to expect that
$\theta (\alpha )$ should be linear in $1\over \alpha$.
Our enumerations indicate that this expected linear
relation does not hold.

The plan of this paper is as follows. In section 2 we
present the natural formulation of conformal invariance
for randomly branched polymers, in section 3 we show
that this formulation leads to inconsistencies, in section 4
we show that the free Parisi-Sourlas theory is
not conformally invariant, in section 5 we
present our results for the exponent $\theta (\alpha )$
obtained by exact
ennumerations of branched polymers in the
cone, and in section 6 we restate our conclusions.

{\bf II. Natural Formulation of Conformal Invariance}

In this section we present the most natural
formulation of conformal invariance for randomly
branched polymers.
The effective Hamiltonian of the field theory
that describes the universal properties of randomly branched polymers\refto{
lubensky} can be written\refto{parisi}
$$
H_{B.P}=\int d^dr[\omega (-{\nabla }^2\phi +V^{\prime}(\phi ) )+
{1\over 2}\omega^2+\bar{\psi}(-
{\nabla }^2 + V^{\prime\prime}(\phi ))\psi ].\eqno(1)
$$
where $V(\phi )={1\over 2}r\phi ^2 +i{g\over 3}\phi^3$.
$\phi$ and $\omega$ are commuting scalars; $\psi$ and $\bar{\psi}$
anti-commute. $H_{B.P.}$
is invariant under the supersymmetry transformations
$\delta\phi =-a\epsilon_{\mu}x^{\mu}\psi$, $ \delta\omega =
2a\epsilon_{\mu}\partial^{\mu}\psi$,
$\delta\psi = 0$, and $\delta\bar{\psi}=a(\epsilon_{\mu}x^{\mu}\omega
-2\epsilon_{\mu}\partial^{\mu}\phi )$
where $a$ is an anti-commuting number and $\epsilon_{\mu}$ an arbitrary
vector\refto{sourlas}.

The supersymmetry imposes relations between the
the correlation functions of the theory. For our
purposes, the following idenities are useful
$$
\cor{\phi (r)\omega (0)}=\cor{\psi (r)\bar{\psi} (0)}=4{\partial\over \partial
r^2}\cor{\phi (r)\phi (0)}.\eqno(2)
$$
{}From these equations
it follows that the the scaling dimensions
$x_{\phi}$, $x_{\omega}$, $x_{\psi}$, and
$x_{\bar{\psi}}$
of the fields $\phi$, $\omega$, $\psi$, and $\bar{\psi}$ are related by
$$
x_{\phi}=x_{\omega}-2=x_{\psi}-1=x_{\bar{\psi}}-1.\eqno(3)
$$
$\phi$ is therefore the lowest dimension field in $H_{B.P.}$.

As stated in the introduction, the most natural formulation of
conformal invariance for randomly branched polymers follows
from the supposition that the most relevant
fields in the randomly branched polymer problem appear
in the effective Hamiltonian eq. 1. If this is
the case, and it would be quite unusual if it were not,
then it follows, by the argument given in the
introduction, that
the lowest dimension field, $\phi$, is primary.

Given that $\phi$ is primary, it follows
from equation ($\call{2}$) that
$$
\omega =k{\nabla }^2\phi +\bar{\omega}\eqno(4)
$$
where $\bar{\omega}$ is a sum of primary operators, and
the constant $k=-{1\over x_{\phi}}$.
This is because in a conformal field theory the two point function of
operators belonging to different conformal families is always zero;
since
$\cor{\phi\omega}$ is not zero, $\omega$ must have a term that belongs
to the conformal family of $\phi$. Since the scaling dimension of $\omega$ is
$x_{\phi}+2$, this term must be a descendent of $\phi$ at level two, and the
only
descendent of $\phi$ at level two which is also a scalar is $\nabla^2\phi$.
Likewise, $\bar\omega$ must be the sum of primary operators, since
if $\bar\omega$ has a term which is not primary, that term
must be the descendent of another operator with scaling
dimension $x_{\omega }-n$ where $n$ is
a positive integer. But again there is no such operator
in the effective Hamiltonian and therefore, in accordance with
our assumptions,
none in the theory (other than $\phi$; $\bar\omega$ is
a commuting variable and therefore cannot be a first
generation descendent of
$\psi$ or $\bar\psi$).
It is important to note that
conformal invariance together with eq. 2 imply that
  $\omega = k\nabla^2\phi +\bar{\omega}$
holds as an operator identity, and not just as
an equality in the plane.

The statement that $\phi$ is primary, together with eq. 4,
constitute our formulation of conformal invariance for
branched polymers.

{\bf III. Consequences of Conformal Invariance}

We now show that this
natural formulation of conformal
invariance for randomly branched polymers leads to inconsistencies.

The correlation function $\cor{\phi (r)\omega (0)}$ is the generating
function for
the number $w_N(r)$ of randomly branched polymer configurations
with $N$ bonds containing the sites
$0$ and $r$\refto{shapir,miller}
$$
\cor{\phi (r)\omega (0)}\sim \int_0^{\infty} dN e^{-\epsilon
N}K_c^Nw_N(r).\eqno(5)
$$
Here $K$ is a fugacity for the number of bonds in the polymer,
$K_c$ is the (non-universal) value
of the fugacity at which the average polymer size
in the fixed fugacity ensemble diverges,
and $\epsilon$ measures the deviations of $K$ from $K_c$:
$\epsilon={(K_c-K)\over K_c}$. Equation ($\call{5}$)
holds for $\epsilon$ small, and $\sim$ means that both sides of the equation
have the same leading
singular behavior.
In the scaling region we expect
$\cor{\phi (r)\phi (0)}={f(r/\xi )\over r^{2x_{\phi}}}$, where
the correlation length $\xi\sim\epsilon^{-\nu}$.
{}From equation ($\call{2}$), $\cor{\phi (r)\omega (0)}={g(r/\xi )\over
r^{2x_{\phi}
+2}}$.
Using this scaling form and integrating both sides of equation ($\call{5}$)
with respect to $r$
and then taking the inverse Laplace transform with respect to $\epsilon$
we have
$$
T^{unrooted}_N\sim\lambda^NN^{-\theta}\sim({1\over K_c})^NN^{\nu
(d-2x_{\phi}-2)-3}\eqno(6)
$$
where $T^{unrooted}_N$ is the number of unrooted branched polymers of size $N$
in the plane, and $\sum_r w_N(r)=N^2T^{unrooted}_N$. Setting $d=2$ and using
the approximate value of $\nu =.64$\refto{derrida}
 and the exact value of $\theta =1$
\refto{parisi},
we find $x_{\phi}=-1/{\nu}\cong -1.5$.  The field $\phi$ therefore has negative
scaling dimension.

Using our proposal for conformal invariance, we can also calculate the
exponent $\theta(\alpha)$, defined by
$T_N(\alpha )\sim \lambda^NN^{-\theta (\alpha )+1}$,
where $T_N(\alpha )$ is the the total number
of randomly branched polymer
configurations rooted near the
apex of a cone with apex angle $\alpha$.\footnote{$^d$}{
A cone can be thought of
as a wedge in the plane, with opposite sides identified.
By apex angle we mean the angle in the plane between the
two sides of the wedge.}
To calculate $\theta (\alpha )$
we need to know
$\cor{\phi\omega}$ in the cone. The transformation $z\to \zeta=z^{{\alpha\over
2\pi}}$ maps the plane onto the cone.
Since, by assumption, $\phi$ is primary, we have at criticality
$$
\cor{\phi (z_1)\phi (z_2)}_{plane}=|{d\zeta\over dz}(z_1)|^{x_{\phi}}
|{d\zeta\over dz}
(z_2)|^{x_{\phi}}\cor{\phi (\zeta_1)\phi (\zeta_2)}_{cone}\eqno(7)
$$
so that,
$$
\cor{\phi (\zeta_1)\phi (\zeta_2)}_{cone}=({2\pi\over\alpha})^{2x_{\phi}}
\mag{\zeta_1}^{({2\pi\over\alpha}-1)x_{\phi}}\mag{\zeta_2}^{({2\pi\over\alpha}-1)x_{\phi}}
{1\over\mag{\zeta_1^{{2\pi\over\alpha}}-\zeta_2^{{2\pi\over\alpha}}}^{2x_{\phi}}}\eqno(8)
$$
where we have normalized $\phi$ so that $\cor{\phi (r)\phi
(0)}_{plane}={1\over\mag{
r}^{2x_{\phi}}}$.
{}From eq. 4 it then follows that
$$
\cor{\phi (\zeta_1)\omega (\zeta_2)}_{cone}=\cor{\phi (\zeta_1)(k\nabla^2\phi
+\bar{\omega})}_{cone}=k\nabla^2_{\zeta_2}\cor{\phi (\zeta_1)\phi
(\zeta_2)}_{cone}\eqno(9)
$$
where we used the orthogonality of $\phi$ and $\bar\omega$
to obtain the second equality. Combining eqs. 8 and 9
then gives
$$
\cor{\phi (\zeta_1)\omega (\zeta_2)}_{cone}=
k\nabla^2_{\zeta_2}\biggl(({2\pi\over\alpha})^{2x_{\phi}}
\mag{\zeta_1}^{({2\pi\over\alpha}-1)x_{\phi}}\mag{\zeta_2}^{({2\pi\over\alpha}-1)x_{\phi}}
{1\over\mag{\zeta_1^{{2\pi\over\alpha}}-\zeta_2^{{2\pi\over\alpha}}}^{2x_{\phi}}}\biggr).\eqno(10)
$$

The correlation function $\cor{\phi (\zeta_1)\omega (\zeta_2)}_{cone}$ is the
the generating function for the total number of branched polymers in the cone.
This
statement is true regardless of whether we put $\phi$ in the bulk and $\omega$
near the apex
or $\omega$ near the apex and $\phi$ in the bulk. On the other hand, we see
from equation ($\call{10}$)
that the exponent $\theta (\alpha )$ depends on whether $\phi$ or $\omega$
is near the apex.
For if we put $\omega$ in the bulk, that is, if we integrate over $\zeta_2$,
holding $\zeta_1$ fixed
and near the apex of the cone, we find
$$
\theta (\alpha )=2-{2\pi\over {\alpha}}\eqno(11)
$$
while if integrate over $\zeta_1$ holding $\zeta_2$ fixed we find something
different,
namely
$$
\theta (\alpha )=2 -{2\pi\over {\alpha}} -2\nu\eqno(12)
$$
for $\alpha <2\pi$.
Since we obtain different answers for $\theta (\alpha )$ depending on whether
we put $\phi$ or $\omega$ near the apex,
equation ($\call{10}$) cannot be correct: the correlation function
$\cor{\phi\omega}$ in the
cone is not equal to the conformal transformation of $\cor{\phi\omega}$ in the
plane, under
the assumption that $\phi$ transforms as a primary operator.
There are in fact two additional problems with equations ($\call{11}$) and
($\call{12}$): (1) the
the coefficient
of ${2\pi \over \alpha}$ is negative, implying that the smaller
the apex angle, the greater the number of rooted polymer configurations, which
is impossible, and (2)
the exponent $\theta (\alpha )$ in equation ($\call{12}$) jumps
discontinuously
from one at $\alpha =2\pi$ to about ${1\over 3}$ for $\alpha$ just smaller than
$2\pi$.
Such a jump seems unlikely, and in any case is in the wrong direction
($\theta(\alpha )$ should
increase as $\alpha $ decreases).

A final point must be considered, namely that the cone is not conformally
related
to the plane, but to the punctured plane.
In most theories, the distinction between the plane and the punctured plane
is irrelevant. A puncture in the continuum theory corresponds to a finite size
hole in the lattice theory. The presence of the hole changes the interactions
between sites variables, and so corresponds to the presence of an energy like
operator. The
energy operator on the lattice can be written as a sum of operators
in the continuum theory. In general, one expects every operator in the
continuum
theory to appear in this sum unless it is forbidden to do so by some symmetry.
The lowest dimension operator in the sum is the most relevant, and dominates
the large distance physics. In most theories the lowest dimension operator
with the same symmetry as the energy operator is the identity operator, so
a defect in the lattice does not change the large distance behavior of
correlation functions. But in the theory we are considering (and in the
Yang-Lee theory) the operator $\phi$ has the same symmetry as the energy
operator
and has a smaller scaling dimension than the identity operator. If, therefore,
$\phi$
appears in the sum (and there is no symmetry which forbids it from doing so)
the large distance behavior of correlation functions on a lattice with a single
defect would be different from the behavior of correlation functions on the
perfect lattice. In the continuum this means that correlation functions in the
plane would not be the same as those in the punctured plane. The exponent
$\theta$ in the punctured plane,
or on a lattice with a finite size
hole, would therefore be different from $\theta$ in the plane.
Since correlation functions in the cone are conformally related to those in the
punctured plane,
they would also be modified, as would predictions for $\theta (\alpha )$.

This proposal, while
raising interesting questions, has a number of
problems.
First,
it is difficult to see how the presence of a finite size
hole on the lattice could alter the exponent $\theta$.
To check this, we have
enumerated all branched polymer
configurations (rooted near the origin) with twelve
and fewer bonds on a square lattice with the
site at the origin removed. Branched polymer
configurations containing the origin were disallowed.
The ratio of the number of polymer configurations
in the presence of the puncture to the number in the plane
appears to converge to a constant near .4 as the number
of bonds in the polymer increases. This indicates that
that the defect does not change $\theta$.
Second, we expect that
the presence of the puncture should not affect the boundary conditions at
infinity. Thus the state at infinity should still be the $SL(2,C)$ invariant
vacuum.  In the Hilbert space formulation, the state corresponding to the
operator $\phi$ is an energy eigenstate, and is orthogonal to the $SL(2,C)$
invariant
vacuum. Hence the partition function $Z$ should not be affected by the
puncture.
Correlation functions, however, will now have an extra $\phi (0)$ inserted in
them.
Since we are interested in two point functions, the correlation functions
with the extra insertion of $\phi (0)$ are three point functions and are
computable. Performing these computations, one can show explicitly that
the correlation function $\cor{\phi\omega}$ in the punctured plane still
depends on whether $\phi$ or $\omega$ is put at the origin.

{\bf IV. Free Parisi-Sourlas Theory Not Conformally Invariant}

In order to gain insight into what might go
wrong with conformal invariance in the branched
polymer theory, we consider here the
simplest theory with a Parisi-Sourlas supersymmetry.
This is obtained by setting the potential
$V(\phi )$ in equation ($\call{1}$)
equal to zero. The commuting fields $\phi$ and $\omega$
then decouple from the anti-commuting fields $\psi$ and $\bar\psi$. Moreover,
the fermion part of the free Hamiltonian $H_F$
has the form of an ordinary Gaussian model,
and is conformally invariant with $\psi$ and
$\bar{\psi}$ transforming as primary fields with conformal dimension zero.
The remaining part of the free theory is
$$
H_F=\int d^dx[\omega (-{\nabla }^2\phi )+{1\over 2}\omega^2].\eqno(13)
$$
or, integrating out the field $\omega$,
$$
H_F=\int d^2r (\nabla^2\phi )^2\eqno(14)
$$
a higher derivative Gaussian theory.
For $H_F$ to be scale invariant,
$\phi$ must have scaling dimension $-1$, or conformal dimension $-1/2$.
Let us assume, as we did for the branched polymers,
 that $\phi$ transforms under conformal transformations as a primary operator.
Then, under the
transformation $z\to z+\epsilon (z)$,
$$
\phi (z,\bar{z})\to (1-{d\epsilon\over dz})^{-{1\over 2}}
(1-{d\bar\epsilon\over d\bar z})^{-{1\over 2}}\phi (z,\bar{z}).\eqno(15)
$$
Taking into account the transformation of the measure $d^2r$ and the
derivatives $\partial_{\mu}$, we find the
variation in $H_F$
$$
\delta H_F=\int d^2r [{d^2\epsilon(z)\over dz^2}
(\bar{\partial}\phi )(\partial\bar{\partial}\phi )+\rm{c.c.}].\eqno(16)
$$
If $\epsilon (z)$ is constant, corresponding to a translation, or linear in
$z$, corresponding to
a dilatation or rotation, $\delta H_F$ vanishes identically, while if $\epsilon
(z)$ is
quadratic in $z$, the integrand is a total derivative, and so $\delta H_F$
vanishes in this case as well.
These three transformations generate the special conformal group, $SL(2,C)$.
On the other hand, the integrand is not a total derivative
for more general conformal transformations.
Thus, assuming that $\phi$ is primary,
the action in equation ($\call{14}$) is invariant under the group $SL(2,C)$,
transformations mapping the plane to itself, but not
invariant under the full conformal group in two dimensions.
Furthermore, there is no way to transform $\phi$ so that $\delta H_F=0$.
To see
this, suppose that under $z\to z+\epsilon (z)$
$$
\phi (z,\bar{z})\to (1-{d\epsilon\over dz})^{-{1\over 2}}
(1-{d\bar\epsilon\over d\bar z})^{-{1\over 2}}
\phi (z,\bar{z})+\delta\phi (z,\bar{z}).\eqno(17)
$$
with $\delta\phi$ linear in $\phi$ and $\epsilon$.
Then
$$
\delta H_F=\int d^2r [({d^2\epsilon(z)\over dz^2}(\bar{\partial}\phi
)(\partial\bar{\partial}\phi ) +
2(\partial\bar{\partial}\phi )(\partial\bar{\partial}\delta\phi ))+
\rm{c.c}].\eqno(18)
$$
In order for $\delta H_F$ to vanish, we need to choose $\delta\phi$ in such a
way that the integrand is
a total derivative.  Considering the first term in equation $\call{18}$ we see
that this is only possible if
$\delta\phi$ is of the form $a\epsilon (z)\partial\phi+b\partial\epsilon
(z)\phi$. But we cannot add a term
of this kind to the transformation law for $\phi$, because this has already
been fixed by translational
and scale invariance
(that is to say, if we add term of this form to
transformation law for $\phi$, $H_F$ will not be
invariant under the
special conformal group).
Thus there is no way to transform $\phi$ so that $\delta H_F=0$ under
general conformal transformations.
Hence the simplest theory with a Parisi-Sourlas supersymmetry is
not conformally invariant even at the classical
level. This may account for the
 problems encountered
in the more complicated theory that describes branched polymers.

{\bf V. Numerical Results}

If randomly branched polymers were conformally invariant,
the form of the transformation law eq.7 for primary operators
leads one to expect a
a linear relationship
between the exponent $\theta (\alpha )$ and
the reciprocal of the cone angle $\alpha$ for
polymers confined to a cone.
A similar relationship would be expected for randomly branched
polymers confined to a wedge but  was not found in a previous analysis of
exact enumeration data\refto{DBL85}.

We have analysed exact enumeration data for lattice trees confined to a
cone. The cone is formed by applying cyclic boundary conditions to a wedge,
cut from a square lattice, in such a way that corresponding lattice sites
on the two boundaries of the wedge become identified as a single site on
the surface of the cone. The number of (weak) embeddings of trees rooted
at the apex of the cone
were enumerated for trees with up to 16 vertices.

We assume that the number of trees on the surface of the cone diverges
as
$$
T_N(\alpha)\sim\lambda^NN^{-\theta(\alpha)+1}\eqno(19)
$$
$\lambda$ is the growth constant for lattice trees on the square lattice
and is the same for trees confined to a wedge or cone as it is for trees
in the bulk lattice\refto{Soteros}. The $y$th moment of the
generating function for $t_N(\alpha)$ will
therefore have critical behaviour described by
$$G^{(y)}(x)=\sum_NT_Nx^NN^y\sim(x_c-x)^{2+y-\theta}\eqno(20)
$$
{}From previous work\refto{DLZtree}, the value of $x_c$ is known to be
$$x_c=1/\lambda=0.19445{{+0.00001}\atop{-0.00002}}.\eqno(21)
$$
To obtain estimates of $\theta(\alpha)$ a Baker-Hunter confluent singularity
analysis\refto{BH} was applied to the exact enumeration data at each of the
cone
angles considered. This type of analysis was used since we expect the presence
of confluences in the generating function for lattice trees\refto{MADRAS}.
The initial analysis used the central estimate of $x_c$ given above
and, since $\theta(\alpha)$ is bigger than unity for the smaller angles,
the first, second and third moments of the generating function
were analysed\refto{DBL85}. The results are
shown in table 1. As this type of analysis may be sensative to the
value of $x_c$ assumed, we also performed an analysis which eliminates
the need to specify $x_c$. To do this the ratio of the number of
embeddings in the bulk lattice $T_N$ and the corresponding number of
embeddings in a cone with wedge angle $\alpha$, $T_N(\alpha)$, was formed
$$
r_N=T_N/T_N(\alpha)\sim N^{\theta(\alpha)-\theta}\eqno(22)
$$
The generating function for these ratios has a critical behaviour
given by
$$
G_r(x)=\sum_Nr_Nx^N\sim(1-x)^{1+\theta(\alpha)-\theta}
.\eqno(22)
$$
This has the advantage that the critical point $(x_c=1)$ is known exactly
and it is not necessary to use higher moments as the exponent is always
greater than 1. The resulting estimates of $\theta(\alpha)$ are shown in
column 4 of table 1.

In most cases the spread in the central estimates of $\theta(\alpha)$
in the columns of table 1, for a given value of $\alpha$, is comparable with
the error estimate in a single column obtained by considering the variation
in the estimate from different Pad\'e approximants to the Baker-Hunter
auxillary function. In order to make a comparison with the expected
linear relation we take the overall estimate of $\theta(\alpha)$ to have
a central value given by an average, over the columns of table 1, of the
central estimates and error bounds such that all of the central estimates
fall within these bounds. These values are plotted against $1/\alpha$ in
figure 1. Even allowing for some degree of subjectivity in the value
of the error bounds the estimates of $\theta(\alpha)$ are clearly not
consistent with a linear relationship. Consequently, although we can not
altogether rule at short series effects as the reason for this deviation from
a linear relation, the results reported here together with the earlier
work on trees confined to a wedge lend support to the
conclusion that randomly branched polymers are not conformally
invariant.

{\bf VI. Conclusions}

In this paper we have presented three arguments
against the conformal invariance, at criticality, of the supersymmetric
field theory that describes randomly branched polymers.
We showed first that the most natural formulation of
conformal invariance leads to inconsistencies, then that
the free Parisi-Sourlas Hamiltonian is not even classically
conformally invariant, and finally, from numerical ennumerations,
that $\theta (\alpha )$ is not linear in the inverse
cone angle $\alpha$.
 None of these arguments constitutes a proof; however
taken together they strongly suggest that
randomly branched polymers are not conformally invariant.
This is interesting, because every other geometrical critical
system with which we are familiar is described by a conformal field theory.
The reason for problems with conformal invariance
may be related to non-unitarity of the model.
Since the theory is non-unitary, the Hamiltonian
 in the
Hilbert space formulation of the problem
is not Hermitian, and therefore need not be diagonalizable.
If $H$ is not diagonalizable, the usual
arrangement of operators in conformal field theories in
highest weight representaions of the Virasoro algebra\refto{belavin},
would be impossible.

{\bf ACKNOWLEDGEMENTS}
J.M. thanks John Cardy for suggesting this problem to him, and for many
helpful conversations. He also thanks Mark Goulian
for discussions.
This work was supported , in part, by NSF Grant PHY 86-14185 and
the Natural Sciences and Engineering Research Council of Canada.

\references

\refis{lubensky}Lubensky, T.C. and Issacson, J., {\sl Phys. Rev. A} {\bf 20},
(1979) 2130.

\refis{parisi}Parisi, G. and Sourlas, N. {\sl Phys. Rev. Lett} {\bf 46}, (1981)
871.

\refis{cardy}see e.g. Cardy, J.L., in {\ Fields, Strings and Critical
Phenomena, Les Houches 1988} E. Brezin and J. Zinn-Justin, eds.,
(Elsevier Science Publishers B.V.) 1989.

\refis{duplantier}see Duplantier, B., {\sl Physica A} {\bf 163}
(1990) 158
and De'Bell K. and Lookman T. "Surface Phase Transitions in Polymer Systems"
Rev. Mod. Phys. (To appear January 1993).

\refis{miller}Miller, J.D., {\sl Europhys. Lett.} {\bf 16}, (1991) 623.

\refis{shapir}Shapir, Y., {\sl Phys. Rev. A.} {\bf 28}, (1983) 1893.

\refis{belavin}Belavin, A.A., Polyakov, A.M., and Zamolodchikov, A.B., {\sl
Nuc. Phys} {\bf B241} (1984), 333.

\refis{sourlas}Parisi, G. and Sourlas, N., {\sl Phys. Rev. Lett.}{\bf 43},
(1979) 744.

\refis{polyakov}Polyakov, A.M., {\sl Sov. Phys. JETP Lett.} {\bf 12},
(1970) 381.

\refis{zinn} Jean Zinn-Justin, private communication.

\refis{derrida}Derrida, B. and Stauffer, D., {\sl J. Phys.} {\bf 46},
(1985) 1623.

\refis{DBL85} De'Bell K. and Lookman T. {\sl Phys.Lett.A} {\bf112A} (1985) 453.

\refis{Soteros} Whittington S.G. and Soteros C. in {\sl Disorder in Physical
Systems} edited by G.R.Grimmett and D.Walsh (Oxford University Press 1990).

\refis{DLZtree} Lookman T., Zhao D. and De'Bell K. {\sl Phys.Rev.A} (1991)
4814.

\refis{BH} Baker G.A. Jr. and Hunter D.L. {\sl Phys.Rev.B} {\bf7} (1973) 3377.

\refis{MADRAS} van Rensburg E.J.J. and N.Madras {\sl J.Phys.A} {\bf25} (1992)
303.

\endreferences

\vfill\eject
{\bf TABLE 1} Estimates of $\theta(\alpha)$ for various cone angles $\alpha$.
Columns labelled 1, 2 and 3 tabulate results obtained from the first,
second and third moment of the generating function respectively. The column
labelled 4 tabulates the results obtained by taking the ratio of the
number of trees in the bulk lattice to the number on the cone.
\settabs\+ccccccccccc&\quad cccccccccc&\quad cccccccccc&\quad cccccccccc&\quad
cccccccccc\cr
\+Angle&1&2&3&4\cr
\+$90^o$&2.091&2.134&2.126&2.114\cr
\+&$\pm0.017$&$\pm0.025$&$\pm0.017$&$\pm0.014$\cr
\+$127^o$&1.98&1.96&1.85&1.852\cr
\+&$\pm0.11$&$\pm0.09$&$\pm0.03$&$\pm0.007$\cr
\+$143^o$&1.85&1.83&1.78&1.73\cr
\+&$\pm0.06$&$\pm0.05$&$\pm0.10$&$\pm0.09$\cr
\+$180^o$&1.55&1.551&1.550&1.553\cr
\+&$\pm0.02$&$\pm0.005$&${{+0.005}\atop{-0.014}}$&$\pm0.010$\cr
\+$233^o$&1.418&1.4080&1.39&1.39\cr
\+&$\pm0.010$&$\pm0.0007$&$\pm0.09$&$\pm0.04$\cr
\+$270^o$&1.2539&1.250&1.252&1.236\cr
\+&$\pm0.0012$&$\pm0.006$&$\pm0.004$&$\pm0.014$\cr
\vfill\eject
{\bf FIGURE 1} Overall estimates of $\theta(\alpha)$ plotted against
$1/\alpha$. The bulk value of $\theta=1$ has been included as the
value for $\alpha=360^o$.

\endit